\documentclass[12]{iopart}
\usepackage{epsfig,amssymb,subfigure,float}
\begin{document}
\title[Transmittivity of a Bose-Einstein condensate
on a lattice: ...]{Transmittivity of a Bose-Einstein condensate
on a lattice: interference from period doubling and the effect
of disorder}
\author{P. Vignolo$^1$, Z. Akdeniz$^{1,2}$ and M.P. Tosi$^1$}
\address{
$^1$ NEST-INFM and Classe di Scienze,
Scuola Normale Superiore,\\I-56126 Pisa, Italy\\
$^2$ Departement of Physics, University of Istanbul, Istanbul, Turkey.}

\begin{abstract}
We evaluate the particle current flowing 
in steady state through a Bose-Einstein condensate 
subject to a constant force in a quasi-onedimensional
lattice and to attractive interactions from fermionic atoms 
that are localized in various configurations inside the lattice wells.
The system is treated within a Bose-Hubbard tight binding model by
an out-of-equilibrium Green's function approach.
A new band gap opens up when the lattice period is doubled by
locating the fermions in alternate wells and yields
an interference pattern in the transmittivity on varying the intensity of
the driving force. 
The positions of the transmittivity minima are determined by matching
the period of Bloch oscillations and the time for tunnelling across 
the band gap.
Massive disorder
in the distribution of the fermions will wash out the interference
pattern, but the same period doubling of the lattice can be
experimentally realized in a four-beam set-up.
We report illustrative numerical results for a mixture of $^{87}$Rb
and $^{40}$K atoms in an optical lattice created by
laser beams with a wavelength of 763 nm.
\end{abstract}
\maketitle

\section{Introduction}
A quasi-onedimensional (1D) optical lattice created by the interference of
two counterpropagating laser beams and superposed
on a highly elongated magnetic trap 
provides an almost ideal periodic
potential for an atomic gas. 
A mixture of a Bose-Einstein condensate and a degenerate fermion gas is an 
interesting system to study in such a confinement.
The interactions can result in the spatial phase
separation of the two components ~\cite{phase-sep} or in the collapse 
of the fermionic component~\cite{collapse}
and generally induce strong changes in the atomic dynamics~\cite{dynamic}, 
depending on the values
of the boson-boson and boson-fermion coupling constants.

A number of experiments have examined properties of ultracold bosonic
atoms and condensates in optical lattices:
Bloch oscillations~\cite{bloch,anderson}, 
Landau-Zener tunnelling~\cite{L-Z}, 
squeezed states~\cite{orzel}, Josephson-like oscillations~\cite{cataliotti},
a superfluid-to-Mott insulator transition~\cite{mott}, 
and 1D band structure~\cite{dalfovo} have all been observed. 
Theoretically, these effects have been studied
by a variety of approaches such as the Gross-Pitaevski 
equation~\cite{GPE},
the discretised non-linear Schr\"odinger equation~\cite{smerzi},
the Bogoliubov-de Gennes equations~\cite{BDG}, the direct diagonalization
of the Bose-Hubbard Hamiltonian~\cite{jaksch}, and quantum MonteCarlo
simulations~\cite{QMC}.

In this work we study the transport properties of the bosons in 
a boson-fermion mixture at zero temperature
under the action of a constant driving force inside a 1D lattice.
A constant acceleration can be imparted to the boson condensate
by using a frequency-chirped optical lattice~\cite{potting}.
For a suitable choice of the system parameters the tunnelling
of the fermionic atoms across the lattice can be neglected and
their presence simply modifies the well depths seen by the bosons.
The interesting cases are those in which the fermionic component
introduces additional periodicities or generates disorder in the lattice
potential. 
All these modifications in the lattice potential could be
more easily realized in the laboratory
by using extra laser beams and/or speckle~\cite{damsky}.
However, we choose to focus on more sofisticated experimental
situations where 
the additional periodicity or the disorder is induced by the fermions,
with the aim to study the magnitude of these effects in such a system
and to address our numerical results to experimental groups who
investigate quantum transport in boson-fermion mixtures.

We describe the system by means of a 1D Bose-Hubbard Hamiltonian,
using a renormalization of the scattering lengths to embody the 
transverse confinement~\cite{salasnich}.
The Hamiltonian is not explicitly solved, but 
the whole lattice is reduced to a single dimer by a renormalization/decimation
procedure~\cite{grosso_rin}. 
Incoming and outgoing leads are connected to
the effective dimer and the steady-state transport coefficient
of the bosons
is inferred from the evaluation of the scattered
wave function of the leads
due to the presence of the dimer.

The paper is organized as follows.
In Sec. \ref{themodel} we present 
the physical system of specific interest 
and the model that we adopt to evaluate
on-site and hopping energies of the 1D Bose-Hubbard Hamiltonian.
In Sec. \ref{trassec} we introduce a scattering matrix formalism
for the case of out-of-equilibrium leads, with the aim of calculating
the bosonic transmittivity through the system. The particle current
is obtained from an extrapolation using a Landauer-like formula.
The method is applied for two choices of the number of fermions,
in cases when they are either regularly 
or randomly distributed inside the lattice. The numerical
results and their physical interpretation  are discussed in 
Sec.~\ref{theresult},
and the main conclusions are given in Sec.~\ref{concl}.
Two Appendices give technical details on the evaluation of
some parameters entering the effective 1D Hamiltonian and on the
recursive algorithm for the calculation of the scattering matrix,
while a third Appendix discusses how period doubling for a pure
Bose condensate can be achieved
in a four-beam experimental set-up.
\section{The model}
\label{themodel}
We focus on a mixture of $^{87}$Rb and $^{40}$K atoms,
which has the peculiarity 
of having a large and negative boson-fermion scattering length.
Let us consider the case in which the number $N_f$ of (spin-polarized)
fermions is much less
than the number $N_b$ of bosons and at most equal to
the number $n_s$ of available lattice sites. 
With a suitable choice of the laser frequency $\omega_L$,
the lattice potential can become considerably deeper for the $^{40}$K 
atoms than for the 
$^{87}$Rb atoms, 
so that the fermions can be taken as localized inside the lattice
potential wells. 

The lattice potential for the two atomic species can be written as
\begin{equation}
U_{\rm K,Rb}(z)=U_{\rm K,Rb}^0\sin^2(kz)
\simeq-
\frac{\hbar\Omega_{\rm K,Rb}^2}{4\delta_{\rm K,Rb}}\sin^2(kz)
\label{primaeq}
\end{equation}
where $\delta_{\rm K,Rb}=\omega_L-\omega_{\rm K,Rb}$ is the detuning of
the laser from the two atomic frequencies
 $\omega_{\rm K}$ and $\omega_{\rm Rb}$,
$k=2\pi/\lambda$ is the wave number of the laser light
and $\Omega_{\rm K,Rb}=d_{\rm K,Rb}E_0/\hbar$ is the Rabi frequency, which is 
determined by 
the laser electric field $E_0$ and by the atomic dipole $d_{\rm K,Rb}$.
The period of the optical lattice is $\lambda/2$.
The atomic dipole depends on the natural width $\Gamma_{\rm K,Rb}$
of the transition~\cite{cohen} so that the ratio $U_{\rm K}^0/U_{\rm Rb}^0$
is given by
\begin{equation}
\frac{U_{\rm K}^0}{U_{\rm Rb}^0}=\frac{\Gamma_{\rm K}}{\Gamma_{\rm Rb}}
\frac{\delta_{\rm Rb}}{\delta_{\rm K}}.
\end{equation}
For the D2 lines of $^{40}$K and $^{87}$Rb, which are used in 
current experiments~\cite{inguscio,jin}, the natural widths are almost equal
($\Gamma_{\rm K}\simeq\Gamma_{\rm Rb}\simeq 6\,$MHz), 
while the atomic frequencies 
have values $\omega_{\rm K}=390.83$\,THz and $\omega_{\rm Rb}=384.227$\,THz.
Choosing for example 
a laser wavelength $\lambda=763$\,nm (or $\omega_L=392.88$\,THz) the
ratio between the depths of the two lattice potentials is 
$U_{\rm K}^0/U_{\rm Rb}^0=4.26$,
so that the transport of the fermions can be neglected in a regime where 
tunnelling motions of the bosons are still allowed.

However, the presence of the localized fermions
affects the transmittivity of the bosons 
by modifying the effective potential that the latter see. 
In the case of a negative mutual
scattering length a well containing a fermion becomes deeper for the
bosons inside it, so that these bosons will have
a lower probability of hopping to the next well.

To study the effect of the fermions on the bosonic transport 
as indicated above, we build 
a 1D tight-binding Hamiltonian for the bosons and 
use a Green's function approach to evaluate their transport 
coefficient through the lattice.
The Bose-Hubbard Hamiltonian for the bosons is
\begin{equation}
H_I=\sum_{i=1}^{n_s} 
\left[E_i |\,i\rangle\langle i\,|+\gamma_i(|\,i\rangle
\langle i+1\,|+|\,i+1\rangle
\langle i\,|)\right]\,.
\label{Hamiltonian}
\end{equation}
Here, the parameters $E_i$ and $\gamma_i$ depend on the number
of bosons in the lattice well labelled by the index $i$
and represent site 
energies and hopping energies, respectively.

We proceed to a 1D reduction of the Hamiltonian by introducing the
transverse widths $\sigma_{\perp\,\rm Rb,K}$ of the bosonic and fermionic
wave functions in a cigar-shaped harmonic trap, together with a 1D
condensate wave function $\phi_i(z)$ in the $i-th$ cell and the 1D
fermion density $n_f(z)$. In a tight-binding scheme $\phi_i(z)$ is a Wannier
function for the bosons in the potential $U_{\rm Rb}(z)$ given in
Eq.(\ref{primaeq}) and, according to the early work of Slater~\cite{slater},
can be written as a Gaussian function,
\begin{equation}
\phi_i(z)=\phi_i(0)\exp[-(z-z_i)^2/(2\sigma_{z\,\rm Rb}^2)],
\end{equation}
where $|\phi_i(0)|^2$ is the number of bosons in the lattice well $i$.
The fermion density is similarly written as a superposition of Gaussian
functions localized in a set of lattice wells labelled by an index $i'$,
\begin{equation}
n_f(z)\propto\sum_{i'}\exp{[(z-z_{i'})^2/\sigma^2_{z\,\rm K}]}.
\end{equation}
The determination of the widths $\sigma_{\perp\,\rm Rb,K}$ and
$\sigma_{z\,\rm Rb,K}$ is carried out variationally, 
as described in~\ref{appxa}.

We can now evaluate the parameters entering the effective
Hamiltonian. The site energies are given by
\begin{equation}
\hspace{-2cm}E_i=\int dz\,\phi_i^2(z)\left[
-\frac{\hbar^2\nabla^2}{2m_{\rm Rb}}+U_{\rm Rb}(z)+\frac{1}{2}g_{bb}
|\phi_i(z)|^2
+g_{bf}
n_f(z)-m_{\rm Rb}az+C_{\rm Rb}\right]
\label{siteenergy}
\end{equation}
where $m_{\rm Rb}$ is the $^{87}$Rb mass, 
$a=F/m_{\rm Rb}$ is the acceleration due to a constant external force $F$
acting on the bosons, $C_{\rm Rb}=\hbar^2/(2m_{\rm Rb}
\sigma_{\perp\,\rm Rb}^2)+\frac{1}{2}m_{\rm Rb}\omega_{\perp\,\rm Rb}^2
\sigma_{\perp\,\rm Rb}^2$,
and $\omega_{\perp\,\rm Rb}$ is the radial frequency of the harmonic 
trapping potential. The parameters $g_{bb}$ and $g_{bf}$ are 
the strengths of the 1D boson-boson and boson-fermion 
interactions, which are given by
\begin{equation}
\left\{
\begin{array}{l}
g_{bb}=\frac{\displaystyle4\pi\hbar^2}{\displaystyle m_{\rm Rb}}
\frac{\displaystyle a_{bb}}{\displaystyle 2\pi\sigma_{\perp\,\rm Rb}^2}\\
g_{bf}=\frac{\displaystyle 2\pi\hbar^2}{\displaystyle m_r}\,
\frac{\displaystyle a_{bf}}{\displaystyle \pi
(\sigma_{\perp\,\rm Rb}^2+\sigma_{\perp\,\rm K}^2)}\,\\
\end{array}\right.
\end{equation} 
with $a_{bb}$, $a_{bf}$ the boson-boson and boson-fermion
scattering lengths~\cite{collapse} and $m_r$ the boson-fermion reduced mass.
Consistently with the model, the hopping energies $\gamma_i$ 
are given by
\begin{equation}
\hspace{-2cm}\gamma_i=\int dz\,\phi_i(z)
\left[
-\frac{\hbar^2\nabla^2}{2m_{\rm Rb}}+U_{\rm Rb}(z)+\frac{1}{2}g_{bb}
|\phi_i(z)|^2
+g_{bf}n_f(z)+
C_{\rm Rb}\right]
\phi_{i+1}(z).
\label{hopenergy}
\end{equation}
This completes the determination of the effective 1D Hamiltonian for the 
bosons, that we shall use to evaluate the transmittivity for various 
distributions of the fermions in the lattice wells.

\section{The transmittivity}
\label{trassec}
\begin{figure}[H]
\centering{
\epsfig{file=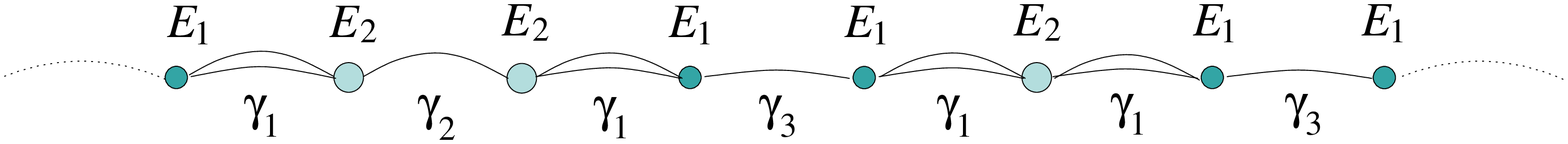,width=1\linewidth}}
\vspace{0.2cm}
\caption{Schematic representation of the 
tight-binding Hamiltonian for a condensate in a 1D lattice, 
with some sites containing a fermion.}
\label{H_tb}
\end{figure}
In Fig.~\ref{H_tb} we show a schematic representation of the Hamiltonian
(\ref{Hamiltonian}) in the absence of an external force. 
For such a system
the boson site energies can be of two types, $E_1$ and $E_2$ 
depending on the absence or presence of a fermion in the well, 
and the boson hopping energies 
can take three values: $\gamma_1$ if the hopping is between a site 
with a fermion and a site without, $\gamma_2$ in the case of two neighboring 
wells which both contain a fermion, and $\gamma_3$ in the case of two 
neighboring wells without fermions. 
We shall consider the following situations: 
({\it i}) the fermionic band
is empty (EF), namely no fermions are present in the lattice; ({\it ii})
the fermionic band is filled (FF), namely there is one fermion per well; 
({\it iii})
the fermionic band is half-filled ($N_f=n_s/2$)
with every other well occupied in an orderly way (OHF); 
and ({\it iv}) the fermionic band is half-filled but the fermions  
are distributed on the lattice in a disordered way (DHF).

In the tight-binding regime the system of bosons behaves in the first two cases as
 a one-band system, which is
described by a single site energy ($E_1$ in the first case and $E_2$ in the 
second) and by a single hopping energy ($\gamma_3$ and $\gamma_2$, 
respectively). The third case describes a two-band boson system with 
two alternating site energies ($E_1$ and $E_2$) and a single hopping energy
($\gamma_1$). 
There is in this case a folding of the dispersion relation in quasi-momentum space,
with the opening of an energy minigap $\Delta E=|E_2-E_1|$
between two sub-bands of equal energy width.
In the fourth case the bosons again form a two-band system of the 
type illustrated in Fig.~\ref{H_tb} and the presence of disorder will cause 
a broadening of the two bands.
We show in~\ref{appxb} how the third case could be realized in 
the laboratory by using a four-beam set-up instead of an ordered
array of fermionic atoms at half occupancy.
The forth case could be realized by adding some speckle to the 
lattice~\cite{damsky}.

\begin{figure}[H]
\centering{
\epsfig{file=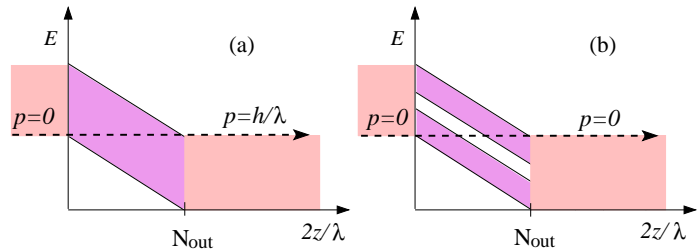,width=0.7\linewidth}}
\caption{Band tilting due to a constant force $F$ and coupling
of the condensate to an incoming and an outgoing lead
for the case of (a) a one-band system and (b) a two-band system.
The position $z$ is in units of $\lambda/2$.}
\label{lead}
\end{figure}

The effect of applying
a constant force to the condensate
can be described as a tilting of the bands in coordinate space,
due to the induced position
dependence of the site energies.
This is shown in Fig.~\ref{lead}:
the situation is analogous to that obtained in an electron gas on a lattice
by applying to it
a static, uniform 
electric field~\cite{zener}. 
The transmittivity of such a system can be calculated, following the 
proposal of Farchioni {\it et al.}~\cite{grosso}, through an evaluation 
of the scattering matrix which explicitly takes into account the presence of 
the bias~\cite{paulsson}. For this purpose, let us connect the system 
to incoming and outgoing leads which mimic its coupling 
to the continuum by injecting and extracting a steady-state particle
current, respectively. The bosons are initially in 
their lowest momentum state ($p=0$) and can leave the lattice 
only at $p=h/\lambda$
(in the one-band case) or at $p=0$ (in a small-minigap two-band
system) if
the coupling to the leads is effected as is shown in Fig.~\ref{lead}.
Elastic Bragg scattering can also occur at both 
these points in quasi-momentum space
as well as at the top of the lower sub-band in the split-band configuration.
As we shall see, in the latter case this produces resonances in the transmittivity.

Let us define $N_{\rm out}$ (with $N_{\rm out}\le n_s$) 
as the number of hops
after which the condensate reaches the highest energy point in
the dispersion curve and
can either leave the lattice or be Bragg reflected.
The two leads have a band width $4\gamma_L$ and their 
centers at $E_L^{\rm in}$ and $E_L^{\rm out}$ are shifted
by $FN_{\rm out}\lambda/2$ in order to 
optimize the coupling with the lattice.
The outgoing wavefunction $\phi_{\rm out}^{\tilde\kappa}(z)$
is an eigenstate of the outgoing lead and
is connected to the incoming wavefunction $\phi_{\rm in}^\kappa(z)$, 
which is an eigenstate of 
the incoming lead, by the relation
\begin{equation}
\phi_{\rm out}^{\tilde\kappa}(z)=\Omega^+(E)\phi_{\rm in}^\kappa(z)\,.
\label{omegapiu}
\end{equation}
The wavevectors
$\kappa$ and $\tilde{\kappa}$ 
of the two wavefunctions are
uniquely identified by the relation
\begin{equation}
\kappa,\tilde{\kappa}=\frac{2}{\lambda}
\arccos\left(\frac{E-E_L^{\rm in,out}}{2\gamma_L}\right).
\end{equation}
The operator $\Omega^+(E)={\Bbb{I}}+G^0(E)T(E)$ in Eq. ({\ref{omegapiu}) involves 
the Green's function $G^0(E)=(E-H_0)^{-1}$ 
of the two leads, 
as well as the scattering matrix $T(E)=H_I[{\Bbb{I}}-G^0(E)H_I]^{-1}$ 
where $H_I$ is the Hamiltonian
acting on the first $N_{\rm out}$
sites of the lattice subject to the force $F$ (see~\ref{appxc}).

In the next step we renormalize
the interaction Hamiltonian
$H_I$ into an effective Hamiltonian $\tilde{H}_I(E)$ acting in the subspace
$\{|\,1\rangle\,,|\,N_{\rm out}\rangle\}$ formed by the two edge sites
that are directly connected to the leads.
In this renormalization approach the 
scattering matrix  
is also projected on the subspace $\{|\,1\rangle\,,|\,N_{\rm out}\rangle\}$ 
and can be written
as a $2\times 2$ matrix and evaluated with a recursive algorithm
as described in~\ref{appxc}.
In such a formalism the transmittivity coefficient ${\cal T}(E)$, which is 
defined as
\begin{equation}
{\cal T}(E)=\frac{\lim_{n\rightarrow+\infty}
\langle n|\phi_{\rm out}^{\tilde\kappa}\rangle
\langle \phi_{\rm out}^{\tilde\kappa}|n\rangle v_{\rm out}}
{\lim_{m\rightarrow-\infty}\langle m|\phi_{\rm in}^\kappa\rangle
\langle \phi_{\rm in}^\kappa|m\rangle v_{\rm in}}
\end{equation}
with $v_{\rm in}$ and $v_{\rm out}$ the velocities of the incoming and
outgoing wavefunction,
is related
to the off-diagonal element $T(E)_{1,N_{\rm out}}$ of the scattering matrix 
by 
\begin{equation}
{\cal T}(E)=4\frac{|T(E)_{1,N_{\rm out}}|^2}{\gamma_L^2}\,
\sin{\left(\kappa\frac{\lambda}{2}\right)}
\sin{\left(\tilde{\kappa}\frac{\lambda}{2}\right)}.
\label{elei}
\end{equation}
It is easy to show that Eq.~(\ref{elei}) is equivalent to the trace of
Eq. (63) in the work of Paulsson~\cite{paulsson}.

The transmittivity determines the particle current passing through the lattice,
and to calculate it we can exploit Landauer's
approach adapted to a bosonic gas with 
out-of-equilibrium leads.
Again following Paulsson~\cite{paulsson}, the current $j$ can be written
\begin{equation}
j=\frac{N_b}{\pi\hbar n_s}\int {\rm d}E
[f(E-\mu_1)-f(E-\mu_2)]{\cal T}(E)
\label{toroo}
\end{equation}
where $f(E-\mu)$ is the Bose-Einstein distribution and $\mu_{1,2}$ 
are the chemical potentials of the reservoirs which 
act as the source and the sink of the bosons.
Note that
the normalization $N_b/n_s$ of the bosonic wavefunction plays the role
of the electric charge in the Landauer formula~\cite{landauer}.
In the zero-temperature limit we have
\begin{equation}
f(E-\mu_1)-f(E-\mu_2)\approx 
\frac{\partial f(E-\mu)}{\partial E}(\mu_2-\mu_1)
\end{equation}
and the density probability
$\partial f(E-\mu)/\partial E$ is a delta function
$\delta(E-\mu)$. Then Eq.~({\ref{toroo}) gives 
\begin{equation}
j=\frac{N_b}{\pi\hbar n_s}{\cal T}(\overline\mu)\Delta\mu
\label{toroo2}
\end{equation}
where $\Delta\mu=\mu_2-\mu_1=N_{\rm out}F\lambda/2$ is the difference in 
chemical potentials and $\overline\mu=(\mu_1+\mu_2)/2$ is their average value,
corresponding to the dashed line in Fig.~\ref{lead}.

The transmittivity may also be expressed in terms of the number of bosons exiting
from the lattice as a drop emitted under a constant driving force, as in the experiment
of Anderson and Kasevich~\cite{anderson}.
The current is written in that context as
\begin{equation}
j=N_{\rm drop}\nu_B,
\label{cold}
\end{equation}
where $\nu_B=(F\lambda)/(2h)$ is the frequency of Bloch oscillations and
$N_{\rm drop}$ is the number of bosons in the first drop. This 
can be calculated
from Eqs.(\ref{toroo2}-\ref{cold}) as
\begin{equation}
\frac{N_{\rm drop}}{N_b}=\frac{N_{\rm out}}{n_s}{\cal T}(\overline\mu).
\label{chiofaluu}
\end{equation}
Equation~(\ref{chiofaluu}) holds for a steady-state current and will not
be applicable to the subsequent drops unless the condensate is continuously
replenished.

\section{Results}
\label{theresult}
We have calculated the transmittivity of the model 
as a function of the acceleration $a$
for the four cases listed at the beginning of Sec.~\ref{trassec}.
The difference $\Delta\mu$ of the 
chemical potentials of the two leads  does not depend on  
$a$ and 
is fixed by the total band width $4\gamma_L$ of the leads, so that 
the current $j$ is directly proportional to the transmittivity 
(see Eq.~(\ref{toroo2})).
We have chosen to focus on the typical experimental parameters used at
LENS~\cite{inguscio} for the $^{87}$Rb-$^{40}$K mixture,
setting $N_b=10^5$, $n_s=200$, $U^0_{\rm Rb}=3.5Er_{\rm Rb}$, and 
$\omega_{\perp\,\rm Rb}=2\pi\times 85.7$\,Hz. As previously remarked,
we set $\lambda=763$\,nm so that we can consider each $^{40}$K atom as
localized in a lattice well.
Moreover, for this choice of the laser wavelength, the detunings 
$\delta_{\rm K,Rb}$ are much larger than the natural widths 
$\Gamma_{\rm K,Rb}$ and spontaneous emission can be neglected.
\begin{figure}[H]
\centering{
\epsfig{file=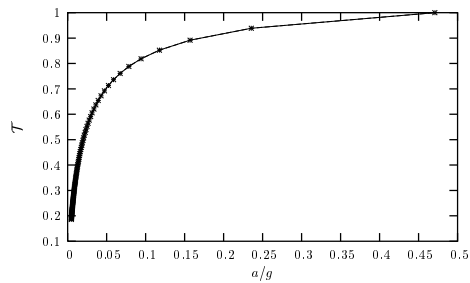,width=0.48\linewidth}}
\caption{The transmittivity as a function of the acceleration $a$ (in units
of the acceleration of gravity $g$) for the cases of an empty fermion band
(EF: +)
and a completely filled fermion band (FF: X).
The lines are guides to the eyes.}
\label{tras_1b}
\end{figure}
In Fig.~\ref{tras_1b} we show the transmittivity of a one-band 
system as a function
of $a$ for the one-band cases in which either the 
fermions are absent
(EF) or each well is occupied by one fermion (FF).
The effect of the presence of the fermions should be that
the boson-fermion attractions diminish the hopping probability 
of the bosons between two neighboring wells,
so that the transmittivity should be lowered. However,
for the case of only one fermion per well and for the usual values
of the boson-fermion scattering length, this effect is not visible
(see Fig.~\ref{tras_1b}).

\begin{figure}[H]
\centering{
\subfigure{
\epsfig{file=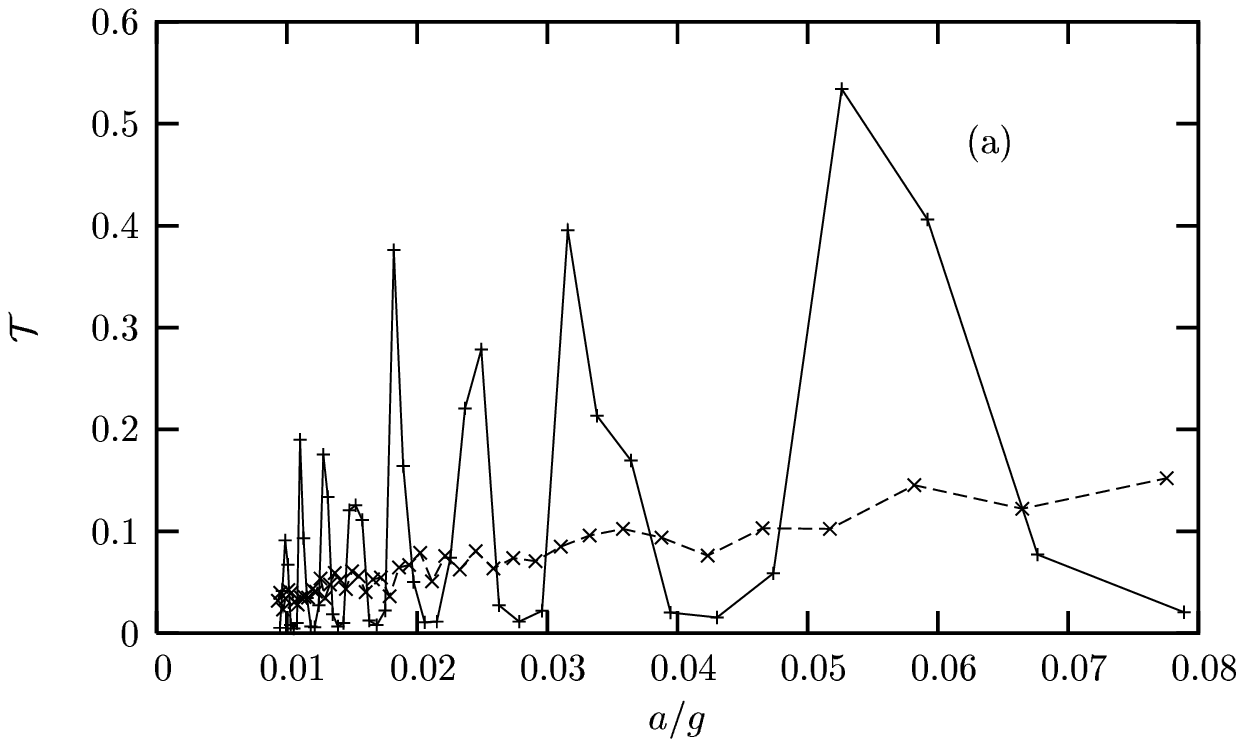,width=0.48\linewidth}}
\subfigure{
\epsfig{file=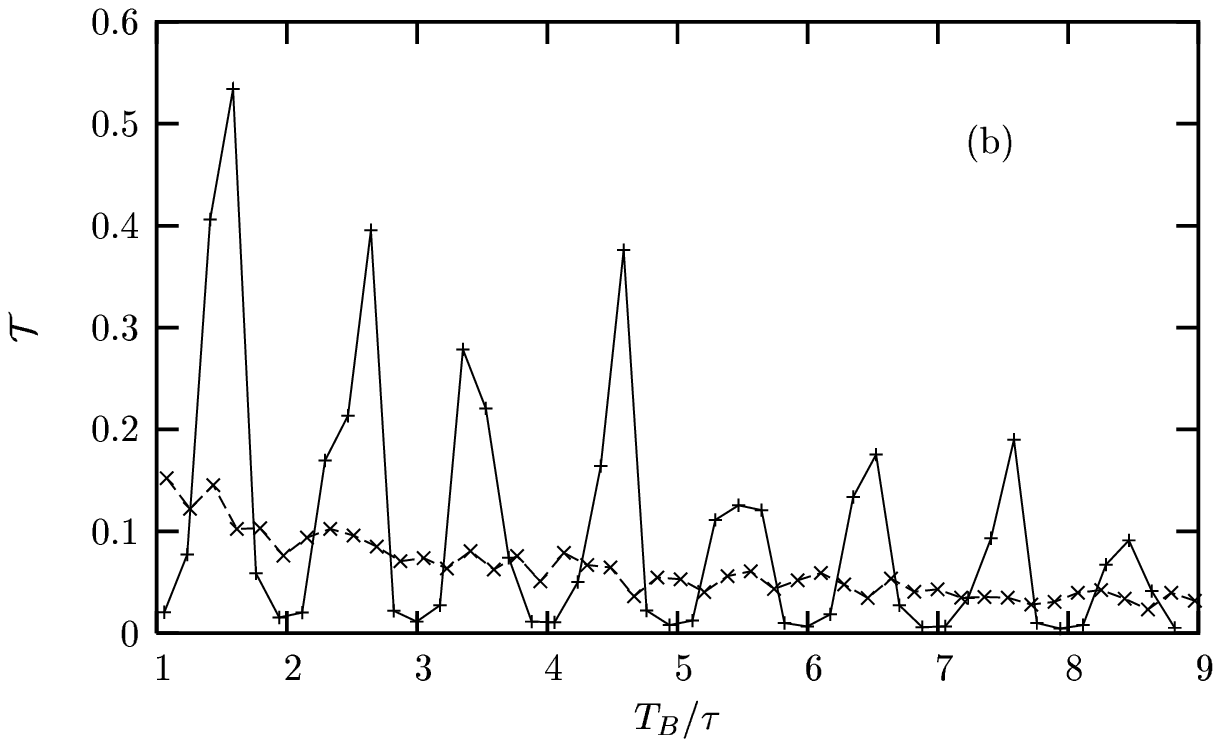,width=0.48\linewidth}}}
\caption{(a) The transmittivity at half occupancy as a 
function of the acceleration $a$ (in units
of the acceleration of gravity $g$) for an ideal two-band system 
(OHF: +) and for randomly distributed fermions (DHF: X).
(b) The same transmittivity data are plotted against
the ratio $T_B/\tau$.
The lines are guides to the eyes.}
\label{tras_2b}
\end{figure}

In Fig.~\ref{tras_2b} we show 
the transmittivity of the two-band system for $N_f=100$ fermions
which are distributed in the lattice wells 
either regularly (OHF) or completely at random (DHF).
The value of $N_{\rm out}$ in these calculations has been taken
to run from 200 to 12: this corresponds to going from left to right
in Fig.~\ref{tras_2b}(a) and from right to left in Fig.~\ref{tras_2b}(b).
The transmittivity in Fig.~\ref{tras_2b} is on average strongly reduced
relative to the cases shown in Fig.~\ref{tras_1b}, but in the OHF case 
it shows a very prominent structure of peaks and troughs.
The observed pattern arises from the interference occuring at $p=-h/2\lambda$
(the left-hand edge of the Brillouin zone in the split-band configuration) 
between the boson wave packets that
are Bragg scattered at $p=h/2\lambda$ in the lower sub-band and those 
that have tunnelled into the upper sub-band, travelled through it,
and tunnelled back into the lower sub-band at $p=-h/2\lambda$.
The two times that govern the interference pattern are 
the Bloch oscillation period 
$T_B=h/(F\lambda)$ and the time $\tau$ taken by the bosons in
coherently tunnelling twice across the minigap, which is proportional
to $(\hbar N_b/\Delta E\,n_s)$~\cite{sakurai}.
We expect constructive interference in the reflectivity at $p=0$
in the upper sub-band, {\it i.e.} destructive interference in the
transmittivity, whenever the ratio $T_B/\tau$ is an integer number.
This behaviour is illustrated in Fig.~\ref{tras_2b}(b) by replotting
the transmittivity data against $T_B/\tau$
with the choice $\tau=(3\pi^2/8)(\hbar N_b/\Delta E\,n_s)$.
Of course this specific value of the proportionality constant
in the tunnelling time, which is consistent with the Heisenberg principle,
depends on the model Hamiltonian that we have assumed.

Returning to Fig.~\ref{tras_2b}(a), the effect of 
disorder in the distribution of
the fermions in the half-occupancy DHF case is to 
strongly diminish the contrast in the interference pattern. 
The disordered distribution that we have
examined in Fig.~\ref{tras_2b} has been produced from a random number 
generator and the pattern is completely washed out in this case of maximum 
disorder. While the average occupancy of a site by a fermion is only
one-half, the transmittivity of the bosons is greatly reduced relative 
to the FF case in Fig.~\ref{tras_1b}. Evidently, the main effect at play
is due to the fluctuations in the disordered distribution of the fermions,
which tend to break the coherence of the Bose condensate.

\section{Concluding remarks}
\label{concl}
In conclusion,
we have studied the condensate 
current induced by a constant external force in 
a 1D optical lattice in the presence of various numbers of fermionic atoms,
distributed over the lattice in orderly or random ways. 
We have focused on the $^{87}$Rb-$^{40}$K mixture, with a choice of the
laser generating the lattice potential 
such that the barriers for the fermions are much higher than 
those for the Bose condensate.
In this regime fermionic transport can be neglected, but 
the presence of the fermions and their
spatial distribution affect
the bosonic transport {\it via} mean-field
attractive interactions.

We have calculated the
transmittivity and the condensate current without
explicitly diagonalizing  the Bose-Hubbard Hamiltonian, but
by exploiting an out-of-equilibrium Green's function approach
adapted to the bosonic case. 
When there is one fermion per well, 
the current is essentially the same as
in the absence of fermions and
increases monotonically with the driving force.
When every second well is occupied by a fermion, instead, the energy spectrum
breaks into two sub-bands and
the probability that the condensate
may tunnel through the band minigap depends on the
energy that it has acquired at the edge of the first sub-band and
on the energy width of the minigap.
Part of the condensate is 
back-scattered when it arrives at the edge of the lower sub-band 
and can interfere at the opposite edge of the Brillouin zone
with the part that has tunnelled
to the upper sub-band and travelled through it.
This gives rise to an interference
pattern showing marked maxima and minima in the transmittivity,
whose location is an integer multiple of a parameter directly proportional
to the energy minigap and inversely
proportional to the driving force.

The interference pattern loses resolution when the distribution of the
fermions in the lattice becomes disordered , up to being completely
washed out in the case of full randomness.
An ordered period doubling of the lattice could be realized, however, with a four-beam experimental set-up.
Observation of resonances in the current as a function of 
the driving force would allow a measurement of the tunnelling time 
through the minigap.

\ack
This work was supported by INFM through the PRA2001-Photonmatter.
ZA also acknowledges support from TUBITAK and from the Research Fund of the
University of Istanbul under Project Number BYP-110/12122002.

\appendix
\section{Determination of radial and axial widths}
\label{appxa}
We adopt the variational method proposed for Bose condensates by
Salasnich {\it et al.}~\cite{salasnich}.
Within a 1D nonpolynomial nonlinear Schr\"odinger equation approach,
we find that the evaluation of the transverse widths 
$\sigma_{\perp\,\rm Rb, K}$ requires the self-consistent solution
of the Euler-Lagrange equations
\begin{equation}
\hspace{-1.5cm}\left\{
\begin{array}{l}
\frac{\displaystyle\hbar^2}{\displaystyle2m_{\rm Rb}
\sigma_{\perp\,\rm Rb}^2}-\frac{\displaystyle1}{\displaystyle2}
m_{\rm Rb}\omega_{\perp\,\rm Rb}^2
\sigma_{\perp\,\rm Rb}^2
+\frac{\displaystyle1}{\displaystyle2}g_{bb}|\phi_i(z)|^2+
g_{bf}\frac{\displaystyle\sigma_{\perp\,\rm Rb}^2}{
\displaystyle\sigma_{\perp\,\rm Rb}^2+
\sigma_{\perp\,\rm K}^2}n_f(z)=0\\
\frac{\displaystyle\hbar^2}{\displaystyle2m_{\rm K}\sigma_{\perp\,\rm K}^2}-
\frac{\displaystyle1}{\displaystyle2}m_{\rm K}\omega_{\perp\,\rm K}^2
\sigma_{\perp\,\rm K}^2+g_{bf}\frac{\displaystyle\sigma_{\perp\,\rm K}^2}
{\displaystyle\sigma_{\perp\,\rm Rb}^2+\sigma_{\perp\,\rm K}^2}|\phi_i(z)|^2=0.
\end{array}\right.
\label{sal}
\end{equation}
Assuming that $\sigma_{\perp\,\rm Rb,K}$ are essentially
constant along the axial position,
we set in Eqs.(\ref{sal}) $|\phi_i(z)|^2\simeq
N_b/(\sqrt{\pi}n_s\sigma_{z\,\rm Rb})$ and
$n_f(z)\simeq 1/(\sqrt{\pi}\sigma_{z\,\rm K})$
or $n_f(z)=0$, depending on the presence or absence of a fermion
in the corresponding well.
The axial widths $\sigma_{z\,\rm Rb,K}$ entering
the axial densities depend themselves
on the radial widths $\sigma_{\perp\,\rm Rb,K}$.
Consequently we first solve Eqs.(\ref{sal}) for given
values of $\sigma_{z\,\rm Rb,K}$, evaluated in the 
absence of interactions and then, having calculated
the radial widths, we evaluate the correction to the axial widths.

Minimization of the bosonic and fermionic
energies with respect to $\sigma_{z\,\rm Rb,K}$ yields
the two coupled equations
\begin{equation}
\hspace{-2.5cm}\left\{
\begin{array}{l}
\frac{\displaystyle\hbar^2}{\displaystyle2m_{\rm Rb}\sigma_{z\,\rm Rb}^2}
-\frac{\displaystyle1}{\displaystyle2}m_{\rm Rb}\omega_{z\,\rm Rb}^2
\sigma_{z\,\rm Rb}^2+
\frac{\displaystyle\hbar^2}{\displaystyle2m_{\rm Rb}\sigma_{z\,\rm Rb}^2}
-\frac{\displaystyle g_{bb}N_b}{\displaystyle 2\sqrt{2\pi}n_s
\sigma_{z\,\rm Rb}}+
\frac{\displaystyle g_{bf}}{\displaystyle\sqrt{\pi}}\,
\frac{\displaystyle\sigma_{z\,\rm Rb}^2}
{\displaystyle(\sigma_{z\,\rm Rb}^2+\sigma_{z\,\rm K}^2)^{3/2}}=0
\\
\frac{\displaystyle\hbar^2}{\displaystyle2m_{\rm K}\sigma_{z\,\rm K}^2}
-\frac{\displaystyle1}\displaystyle{2}m_{\rm K}\omega_{z\,\rm K}^2
\sigma_{z\,\rm K}^2+
\frac{\displaystyle g_{bf}N_b}{\displaystyle\sqrt{\pi}n_s}\,
\frac{\displaystyle\sigma_{z\,\rm K}^2}
{\displaystyle(\sigma_{z\,\rm Rb}^2+\sigma_{z\,\rm K}^2)^{3/2}}=0
\end{array}\right.
\end{equation}
for the case of a well containing a fermion, while in the
absence of a fermion we have
\begin{equation}
\frac{\hbar^2}{2m_{\rm Rb}\sigma_{z\,\rm Rb}^2}
-\frac{1}{2}m_{\rm Rb}\omega_{z\,\rm Rb}^2
\sigma_{z\,\rm Rb}^2+
\frac{g_{bb}N_b}{2\sqrt{2\pi}n_s\sigma_{z\,\rm Rb}}=0
\label{laurea}
\end{equation}
as already found by Chiofalo {\it et al.}~\cite{BDG}.
Here, the axial frequencies $\omega_{z\,\rm Rb,K}$ for a single well
are expressed in terms of
the potential depths $U_{\rm Rb,K}^0$ and of the recoil energies
$Er_{\rm Rb,K}=\hbar^2k^2/2m_{\rm Rb,K}$ as
\begin{equation}
\omega_{z\,\rm Rb,K}=\frac{1}{\hbar}
\sqrt{4\,U^0_{\rm Rb,K}Er_{\rm Rb,K}}.
\end{equation}
The solution of Eqs.~(\ref{sal})-(\ref{laurea}) provides the further input that is needed in Eqs.~(\ref{siteenergy}) and (\ref{hopenergy}).

\section{Period doubling with a four-beam set-up}
\label{appxb}
Four laser beams can be used to create a 1D lattice
with a double periodicity along the $z$ direction.
Two beams are set in a counterpropagating configuration
along the $z$ axis,
while the other two beams
are rotated by angles of 60$^{\circ}$ and 120$^{\circ}$
with respect to the $z$ axis.
In the region of space where the four beams are superposed, the longitudinal
electric field $E$ is proportional to
\begin{equation}
E\propto \beta e^{i\alpha} (e^{ikz/2}+
e^{i\phi_1}e^{-ikz/2})
+(e^{ikz}+e^{i\phi_2}e^{-ikz}).
\label{electricfield}
\end{equation}
In Eq.~(\ref{electricfield}) 
$\beta$ and $\alpha$ are 
the relative amplitude and phase 
of the rotated beams with respect to
the counterpropagating pair, $\phi_1$ is 
the relative phase of the rotated beams, and $\phi_2$
the relative phase of the counterpropagating beams.
All these parameters can be set by choosing suitable optical components.

Upon setting $\phi_1=\phi_2=\pi$ and $\alpha=\pi/2$, the resulting
lattice potential will have the shape
\begin{equation}
U(z)\propto\beta^2\sin^2(kz/2)+\sin^2(kz)
\label{potsite}
\end{equation}
and the double period enters through the well depths, 
that is the site energies, which will differ by a factor $\beta^2$.
Instead, by setting $\phi_1=\phi_2=0$ and $\alpha=\pi/2$
we have
\begin{equation}
U(z)\propto\beta^2\cos^2(kz/2)+\cos^2(kz)
\end{equation}
and the potential hills 
(that is the hopping energies) have alternating values.
In the case of a $^{87}$Rb condensate, a 
lattice potential of the type in Eq. (\ref{potsite}) with $\beta\simeq 0.1$
is equivalent to the effective potential generated by a regular distribution
of $^{40}$K atoms inside every other well of an optical lattice 
created with a single pair of laser beams.

\section{The recursive algorithm} 
\label{appxc}
Figure~\ref{h_rin} shows a schematic representation of 
the whole Hamiltonian for the system
composed of a finite lattice and two leads, and of its decomposition
into the sum of $H_0$ and $H_I$ for the evaluation of the scattering matrix.
We show below how $H_I$ can be reduced to an effective Hamiltonian
$\tilde{H}_I$ for a dimer in a subspace 
$\{|\,1\rangle,|\,N_{\rm out}\rangle\}$ and
how this leads to an expressions for the scattering matrix
and for the outgoing wavefunctions.

\begin{figure}[H]
\centering{
\epsfig{file=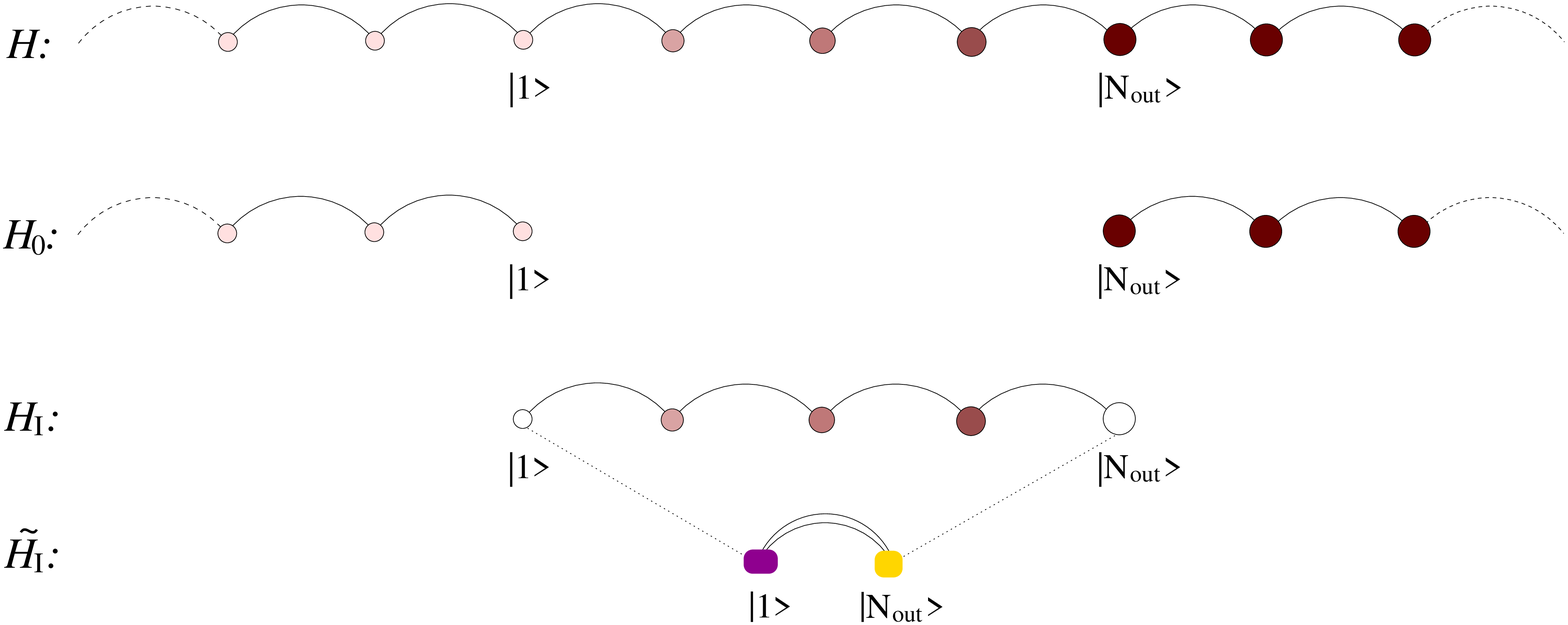,width=1\linewidth}}
\vspace{0.3cm}
\caption{Schematic representation of the Hamiltonian $H=H_0+H_I$
of the system composed by the leads ($H_0$) and by
the lattice ($H_I$), 
and of the effective Hamiltonian $\tilde H_I$ projected on the two edge sites.}
\label{h_rin}
\end{figure}

Let $n$ be the site index number. To calculate the transmitted wavefunction
we have to evaluate 
\begin{equation}
\langle n\,|\,\phi_{\rm out}^{\tilde\kappa}\rangle=
\sum_m\langle n\,|\,{\Bbb I}+G^0(E)T(E)\,|\,m\rangle\langle m\,
|\phi_{\rm in}^\kappa\rangle
\label{anna}
\end{equation}
for $n>N_{\rm out}$.
Since the wavefunctions $\phi_{\rm in}^\kappa$  and 
$\phi_{\rm out}^{\tilde\kappa}$
are defined in two disconnected spaces,
Eq.~(\ref{anna}) becomes
\begin{eqnarray}
\langle n\,|\,\phi_{\rm out}^{\tilde\kappa}\rangle&=&
\sum_{m\le1,l}\langle n\,|\,G^0(E)\,|\,l\rangle\langle l\,
|\,T(E)\,|\,m\rangle\langle \,m|\phi_{\rm in}^\kappa\rangle \nonumber\\
&=&\langle n\,|\,G^0(E)\,|\,N_{\rm out}\rangle\langle N_{\rm out}
\,|\,T(E)\,|\,1\rangle\langle 1\,|\phi_{\rm in}^\kappa\rangle\nonumber\\
&=&\sqrt{2}[G^0(E)]_{n,N_{\rm out}}[T(E)]_{N_{\rm out},1}
\sin(\kappa\lambda/2)
u_{\kappa}(\lambda/2)\,,
\label{anna2}
\end{eqnarray}
where we have set 
$\langle 1\,|\phi_{\rm in}^\kappa\rangle=u_{\kappa}(\lambda/2)
[\exp{(i\kappa \lambda/2)}-
\exp{(-i\kappa \lambda/2)}]/
i\sqrt{2}$ with $u_{\kappa}(\lambda/2)=\phi_i(z)$.
It can be shown
that Eq. (\ref{anna2}) is equivalent to Eq.~(34) 
in the work of Paulsson~\cite{paulsson} 
by making use of the relation $G^0(E)T(E)=G(E)H_I(E)$.

The Green's function element $[G^0(E)]_{n,N_{\rm out}}$ in Eq.(\ref{anna2})
determines the coherence
between site $n$ and site $N_{\rm out}$ on the chain
for the outgoing lead. It can be written as
\begin{equation}
[G^0(E)]_{n,N_{\rm out}}=\frac{\gamma_L^{n-N_{\rm out}}}
{|\gamma_L^{n+1-N_{\rm out}}|}e^{i\tilde{\kappa}(n+1-N_{\rm out})\lambda/2}
\end{equation}
in terms of quantities defined in the main text.
The scattering matrix element $[T(E)]_{N_{\rm out},1}$ can be evaluated
from the effective Hamiltonian $\tilde{H}_I(E)$ as schematically
illustrated in Fig.~\ref{h_rin}:
\begin{equation}
[T(E)]_{N_{\rm out},1}=\langle N_{\rm out}\,|\tilde{H}_I(E)[{\Bbb I}_2-
\tilde{G}^0(E)\tilde{H}_I(E)]^{-1}|\,1\rangle
\label{gola}
\end{equation}
where ${\Bbb I}_2$ is the $2\times 2$ identity matrix and
$\tilde{G}^0(E)$ is the Green's function of the leads projected
onto the subspace $\{|\,1\rangle,|\,N_{\rm out}\rangle\}$.
The only non-zero elements of the $2\times 2$ matrix $\tilde{G}^0(E)$
are the diagonal ones, which are given by
${\cal G}(E-E_L^{\rm in},\gamma_L)$ and 
${\cal G}(E-E_L^{\rm out},\gamma_L)$,
with
\begin{equation}
{\cal G}(E-E_s,\gamma_L)=\frac{1}{2\gamma_L^2}[E-E_s-
\sqrt{(E-E_s)^2-4\gamma^2_L}]\,.
\end{equation}

Finally,
the elements of the matrix $\tilde{H}_I(E)$ in Eq.~(\ref{gola}) can 
be calculated by decimation of
the sites (2, 3, $\dots$, $N_{\rm out} -1$) and can be written
in the form of continued fractions. That is,
\begin{equation}
[\tilde{H}_I(E)]_{1,1}=[H_I]_{1,1}+
\frac{\displaystyle[H_I]_{1,2}^2}{\displaystyle E-[H_I]_{2,2}
-\frac{\displaystyle[H_I]_{2,3}^2}{\displaystyle E-[H_I]_{3,3}-\dots}
}\,\label{diagonal}
\end{equation}
and
\begin{equation}
\hspace{-2.5cm}\protect[\tilde{H}_I(E)]_{1,N_{\rm out}}=
\protect[H_I]_{1,2}\frac{\displaystyle 1}{\displaystyle E-[H_I]_{2,2}}
[H_I]_{2,3}
\frac{\displaystyle1}{\displaystyle E-[H_I]_{3,3}-
\frac{\displaystyle[H_I]_{3,2}^2}{\displaystyle E-[H_I]_{2,2}}}
[H_I]_{3,4}\dots\,,
\end{equation}
with $[\tilde{H}_I(E)]_{N_{\rm out},1}=[\tilde{H}_I(E)]_{1,N_{\rm out}}$
and $[\tilde{H}_I(E)]_{N_{\rm out},N_{\rm out}}$ given similarly to
$[\tilde{H}_I(E)]_{1,1}$ in Eq.~(\ref{diagonal}).

\end{document}